# An Analysis of Indexing and Querying Strategies on Technologically Assisted Review Tasks

Alexandros Ioannidis
University of Strathclyde
Glasgow, UK
alexandros.ioannidis@strath.ac.uk

## ABSTRACT
This paper presents a preliminary experimentation study using the CLEF 2017 eHealth Task 2 collection for evaluating the effectiveness of different indexing methodologies of documents and query parsing techniques. Furthermore, it is an attempt to advance and share the efforts of observing the characteristics and helpfulness of various methodologies for indexing PubMed documents and for different topic parsing techniques to produce queries. For this purpose, my research includes experimentation with different document indexing methodologies, by utilising existing tools, such as the Lucene4IR[1] (L4IR) information retrieval (IR) system, the Technology Assisted Reviews (TAR) for Empirical Medicine tool[2] for parsing topics of the CLEF collection and the TREC evaluation tool[3] to appraise system's performance. The results showed that including a greater number of fields to the PubMed indexer of L4IR is a decisive factor for the retrieval effectiveness of L4IR.

## KEYWORDS
Medical Information Retrieval, PubMed, CLEF Collection, Indexing Schemata, Query Parser, BM25

## 1 INTRODUCTION

The purpose of this paper is to find the best combination of indexing methodology and query parser variation technique to perform for the initial round of retrieval of the CLEF 2017 eHealth Task 2. This work is important because it helps the process of discovering an optimal methodology to minimise the time and cost of reviewing medical documents by medical experts, which is associated with certain stages of TARs, such as retrieval and ranking [4].

The recently entrenched CLEF 2017 eHealth Task 2 required participants to rank a set of PubMed abstracts (A) given the results retrieved from the previous Task 1. The track had 2 objectives, create an effective ranking of the documents, such that all of the relevant abstracts are recovered as soon as possible and determine a subgroup of (A), which contains as many of the relevant abstracts for the least effort [5]. The participants were given the associated qrels judgments file, which follows the TREC format Topic number, Iteration, PubMed identification number, Relevancy, a binary code of 0 (non-relevant) and 1 (relevant). The order of documents in the qrels file does not suggest degree of relevance. Documents not occurring in the qrels file were not judged by the human assessor and are presumed to be irrelevant in the evaluations. A more detailed description of the task can be found here[4].

It is important to examine some of the different approaches to the CLEF 2017 eHealth Task 2 to identify gaps for exploration (see research questions below) and to also ensure I create a realistic baseline for my evaluations. For example, Alharbi et al. [1] utilised the review title and Boolean query to order the abstracts retrieved by the query, by applying standard similarity measures. According to the authors [1] the title and terms extracted from the Boolean query contributed the most useful information for this task. Their methodology made use of 3 topic parts, the text of the title, the terms and Mesh items parsed from the Boolean query. The Mesh items were preprocessed with the same approach applied on the Boolean query. The text was tokenised, changed to lower case, stop words were taken out and the rest of the tokens were stemmed. Lastly, the data extracted from the topic and every abstract was converted into TF-IDF weighted vectors [1] to calculate similarity among the topic and every abstract using the cosine metric for the pair of vectors.

Ecnu et al. [3] used a customised Learning-to-Rank (L2R) model and the word2vec to represent queries and documents and compute their similarities by cosine distance. Their L2R model consists of 3 points, query expansion, feature extraction and model training. In the query expansion stage, they improved retrieval precision by expanding queries. In the feature extraction stage, they extracted features of each query document pair. Finally, in the learning phase of the L2R model, the relevance of a query-document pair was assessed with the random forest classifier. Norman et al. used a system that builds on logistic regression [6], and implemented strategies to handle class imbalance and perform relevance feedback. Additionally, they tested 2 classifiers, logistic regression with stochastic gradient descent learning on the entire data and standard logistic regression trained with conventional methods on a subgroup of the data with added preprocessing to enhance the yield. Nunzio et al. [7] concentrated their work on discovering the optimal combination of certain hyper-parameters by utilising the training data available and a force brute strategy. This strategy created different query features of the identical information need, given a minimum amount of relevance feedback. Furthermore, they decided to use a plain BM25 retrieval model and acquire the relevance feedback for the first abstract in the ranking list for each topic. Afterwards, they asked 2 different people to build 2 different queries according to the value of the feedback. Finally, alternative strategies were designed that use the following parameters, number of documents to assess in batches or iteratively, percent of documents to assess, maximum number of documents to assess per iteration, the minimum precision the system can reach before terminating the search and more.

---

[1] https://github.com/lucene4ir [Accessed: 2018-01-26]
[2] https://github.com/CLEF-TAR/tar [Accessed: 2018-02-02]
[3] http://trec.nist.gov/trec_eval [Accessed: 2018-02-01]

---

[4] ://sites.google.com/site/clefehealth2017/task-2

Singh et al. [8] utilised Lucene's inverted file to index the retrieved articles retrieved from PubMed query performed during a systematic review. The query is processed for term boosting, fuzzy search and used for scoring documents according to TF-IDF similarity. Relevance feedback is used to update the query and become more pragmatic. Essentially, different people have implemented different strategies and they all used different methods, so it is hard to determine what indexing strategy or querying parsing technique is best to perform for the initial round of retrieval.

Research Questions: Therefore, the following research questions are raised: **(RQ1)** How do different indexing schemata affect performance? **(RQ2)** How does the initial query extracted from the CLEF Topic affect performance?

## 2 METHOD

Data and Materials: The data that used is part of the CLEF collection[5]. The CLEF 2017 eHealth Task 2 collection consists of 198,366 PubMed XML documents and it contains 50 topics and queries. Furthermore, each PubMed article set has a structure, which contains the Medline citation fields and the PubMed data and metadata fields. The Medline citation includes fields such as the ID of the PubMed document (PMID), date of creation and completion, ISO abbreviations, title and abstract of the article, the list of author(s), the MeSH heading list, the list of chemicals other information of the Medline Journal. The PubMed data contain information such as historical dates, the publication status and more information.

Each topic of the CLEF 2017 eHealth Task 2 collection contains the topic-ID, the title of the review and the Boolean query manually drafted and constructed by Cochrane experts and the collection of PMIDs. For all of my indexes I used the standard tokeniser in L4IR, and multiple token filters. The token filters used are the fol-lowing, stopping, porter stemmer, generation of word parts, pattern replacement, lowercase and word delimiter with the key parameter of this token filter set to 'preserveOriginal'. Additionally, I utilised L4IR [2], which is a collaborative effort of researchers to extend the Lucene library and produce a group of statistical evaluation tools for IR methods such as indexing and retrieval for different types of test collections.

L4IR, includes among other functions, 2 main applications, the IndexerApp and the RetrievalApp. The IndexerApp, allows indexing of multiple different TREC collections, such as the Aquaint and PubMed collections. The RetrievalApp, is a batch retrieval application, which contains various retrieval algorithms, for example the BM25, the PL2 and more. Both the IndexerApp and the RetrievalApp applications can be configured by editing their parameters files 'index_params.xml' and 'retrieval_params.xml'. For the purposes of this study, I used the BM25 baseline retrieval algorithm used most commonly at CLEF 2017 eHealth Task 2, to rank the document sets. The 1st indexing schema (see section 2.1) was used as a baseline.

Moreover, I utilised the CLEF TAR tool, for extracting parts from the topics of the CLEF 2017 eHealth Task 2 collection, such as the ID, title and query of the topic and more. Additionally, I used Python and Excel to automate the process of running experiments and statistical significance tests, but also to facilitate further analysis in the future, in a larger experimentation setup (the incorporation of a greater number of indexing methodologies and query parsing techniques).

**Measures:** The metrics I used for the purpose of this research are Mean Average Precision (MAP), Precision after 10, 20 and 30 documents and Interpolated Recall - Precision Averages (IRPA) at 0.10, 0.20 and 0.30 recall [9].

### 2.1 Indexing Schemata

In my research I used 5 different indexing schemata, which are outlined below. Each indexing schema consists of a unique combination of XML fields.

**baseline:** where I indexed the Title, Abstract and PMID;
**1+AJY:** where I included the fields of the 1st indexing schema along with the Author(s), the Journal title and the Year;
**2+MHL:** where I combined the XML fields of the 2nd indexing schema with the Mesh Heading List (MHL);
**2+MTA:** where I included the 2nd indexing schema and the MedlineTA (MTA) field; and
**2+MHLMTA:** where I combined the 2nd indexing schema with the Mesh HL and the MedlineTA.

### 2.2 Query Parser Variations

Furthermore, 5 different query parsers were implemented, so that the performance of the IR system L4IR could be tested by using different query variations. The 5 different query parsers that were used are listed below:

**title:** The Title of the topic;
**query:** The extracted Query of the topic;
**title&query:** The combination of the Title and Query of the topic;
**query2:** The alternatively[6] extracted Query of the topic; and
**title&query2:** The combination of the Title and alternative Query of the topic.

**Retrieval Model and Parameters:** As mentioned above the IR model used was BM25 and its free parameters b and k were set to be 0.75 and 1.2 respectively, for the main experimentation round. But I also sweeped over b and k parameter. To compare my results and examine statistical significance, I conducted paired t-tests in a common batch-style experimentation setup [9]. I conducted multiple paired t-tests using the samples of the two runs being assessed each time. Each sample consisted of 50 queries corresponding to the 50 topics of the CLEF 2017 eHealth Task 2 collection. Furthermore, I denoted statistical significance at 5%. I used the * symbol to indicate significance when p-value <= 0.05, the symbol + when p − value < 0.01 and the symbol ++ when p−value < 0.001. In the tables below, I have captured some of the MAP metrics (2) of my retrieved results from L4IR using the trec_eval tool.

## 3 RESULTS

The figures below display the information captured from my experiments in the tables above. In Figure 1 the horizontal axis exhibits the 5 different query parsing variations. Moreover, because the fifth indexing methodology incorporates the most fields of the PubMed

---

[5]http://www.clef-initiative.eu/dataset/test-collection [Accessed: 2018-01-11]

[6]Removal of term explosion and search over Title/Abstract symbols and the replacement of zero/one-character symbols with the PubMed wildcard symbol.

**Table 1: Comparison of the MAP metric between the 5 different indexing schemes and the 5 different query parsers, when b = 0.75 and k = 1.2.**

| Query Parser | TAP | 1+AJY | 2+MHL | 2+MTA | 2+MHLMTA |
|---|---|---|---|---|---|
| title | 0.1211 | 0.1237 | 0.135$^+$ | 0.1233 | 0.136$^+$ |
| query | 0.039 | 0.0419 | 0.048$^*$ | 0.0419 | 0.0485$^+$ |
| title+query | 0.0836 | 0.0954$^*$ | 0.0993$^*$ | 0.0942$^*$ | 0.1013$^{++}$ |
| query2 | 0.0775 | 0.0846 | 0.0906$^+$ | 0.086 | 0.0916$^+$ |
| title+query2 | 0.1025 | 0.1154 | 0.1183$^+$ | 0.1154 | 0.1197$^{++}$ |

**Table 2: Comparison of the MAP metric between the 5 different indexing schemes and the 'title&query2' parser, while sweeping the k parameter and b = 0.75.**

| k | TAP | 1+AJY | 2+MHL | 2+MTA | 2+MHLMTA |
|---|---|---|---|---|---|
| 1.0 | 0.1016 | 0.1146 | 0.1175$^+$ | 0.1139 | 0.1194$^{++}$ |
| 1.2 | 0.1025 | 0.1154 | 0.1183$^+$ | 0.1154 | 0.1197$^{++}$ |
| 1.4 | 0.1034 | 0.116 | 0.1184$^+$ | 0.1161 | 0.1212$^+$ |
| 1.6 | 0.1036 | 0.1137 | 0.1189$^+$ | 0.1145 | 0.1205$^+$ |
| 1.8 | 0.1034 | 0.1116 | 0.121$^+$ | 0.1122 | 0.1219$^{++}$ |
| 2.0 | 0.1048 | 0.1109 | 0.1203$^+$ | 0.1119 | 0.1213$^+$ |

**Table 3: Comparison of the MAP metric between the 5 different indexing schemes and the 'title&query2' parser, while sweeping the b parameter and k = 1.2.**

| b | TAP | 1+AJY | 2+MHL | 2+MTA | 2+MHLMTA |
|---|---|---|---|---|---|
| 0.0 | 0.1102 | 0.1136 | 0.1282$^+$ | 0.1144 | 0.1284$^+$ |
| 0.1 | 0.1102 | 0.1136 | 0.1282$^+$ | 0.1145 | 0.1297$^+$ |
| 0.2 | 0.1115 | 0.1145 | 0.1288$^+$ | 0.1145 | 0.1297$^+$ |
| 0.3 | 0.1105 | 0.1183 | 0.1268$^+$ | 0.1184 | 0.1294$^+$ |
| 0.4 | 0.1101 | 0.1188 | 0.126$^+$ | 0.1187 | 0.1282$^{++}$ |
| 0.5 | 0.1092 | 0.1181 | 0.1243$^+$ | 0.1172 | 0.1258$^+$ |
| 0.6 | 0.1077 | 0.117 | 0.1227$^+$ | 0.1162 | 0.1246+ |
| 0.7 | 0.1036 | 0.116 | 0.1204$^+$ | 0.1161 | 0.1225$^{++}$ |
| 0.8 | 0.1009 | 0.1106 | 0.1158$^+$ | 0.1139 | 0.1178$^{++}$ |
| 0.9 | 0.097 | 0.1073 | 0.1126$^+$ | 0.1071 | 0.1139$^{++}$ |
| 1.0 | 0.092 | 0.1037$^*$ | 0.1067$^+$ | 0.1036$^*$ | 0.1082$^{++}$ |

**Table 4: Comparison of the MAP metric between the 5 different indexing schemata and the 'title&query' parser, while sweeping the k parameter and b = 0.75.**

| k | TAP | 1+AJY | 2+MHL | 2+MTA | 2+MHLMTA |
|---|---|---|---|---|---|
| 1.0 | 0.0838 | 0.0929 | 0.0994$^+$ | 0.0919 | 0.1012$^+$ |
| 1.2 | 0.0836 | 0.0954$^*$ | 0.0993$^+$ | 0.0942 | 0.1013$^+$ |
| 1.4 | 0.0842 | 0.0965$^*$ | 0.0999$^+$ | 0.0957 | 0.1027$^+$ |
| 1.6 | 0.0832 | 0.094 | 0.1001$^+$ | 0.094 | 0.1021$^+$ |
| 1.8 | 0.0822 | 0.0942 | 0.1017$^+$ | 0.0937 | 0.1032$^+$ |
| 2.0 | 0.0806 | 0.0939 | 0.1012$^+$ | 0.0933 | 0.1027$^+$ |

XML documents compared to the other 4 indexing methodologies, it was portrayed with a distinctive dashed line in order to draw attention into its performance against the rest. This allows me to draw easier conclusions on whether adding more XML fields into my indexer increases the performance of my IR system.

**Table 5: Comparison of the MAP metric between the 5 different indexing schemes and the 'title&query' parser, while sweeping the b parameter and k = 1.2.**

| b | TAP | 1+AJY | 2+MHL | 2+MTA | 2+MHLMTA |
|---|---|---|---|---|---|
| 0.0 | 0.0839 | 0.0897 | 0.0998$^+$ | 0.0887 | 0.1007$^+$ |
| 0.1 | 0.0887 | 0.0921 | 0.1019$^*$ | 0.0927 | 0.1027$^{++}$ |
| 0.2 | 0.0902 | 0.0939 | 0.104$^+$ | 0.0938 | 0.1046$^+$ |
| 0.3 | 0.0903 | 0.0955 | 0.1045$^*$ | 0.0952 | 0.1052$^+$ |
| 0.4 | 0.0898 | 0.096 | 0.1053+ | 0.0959 | 0.1065$^+$ |
| 0.5 | 0.0888 | 0.0957 | 0.1042$^+$ | 0.0944 | 0.106$^+$ |
| 0.6 | 0.088 | 0.096 | 0.1028$^+$ | 0.0947 | 0.1048$^+$ |
| 0.7 | 0.0849 | 0.0952$^*$ | 0.101$^+$ | 0.0945 | 0.1032$^{++}$ |
| 0.8 | 0.0826 | 0.0934 | 0.0977$^+$ | 0.0935 | 0.0999$^{++}$ |
| 0.9 | 0.0791 | 0.0903 | 0.0944$^+$ | 0.0897 | 0.0965$^{++}$ |
| 1.0 | 0.0739 | 0.0865$^*$ | 0.0891$^+$ | 0.086 | 0.0916$^+$ |

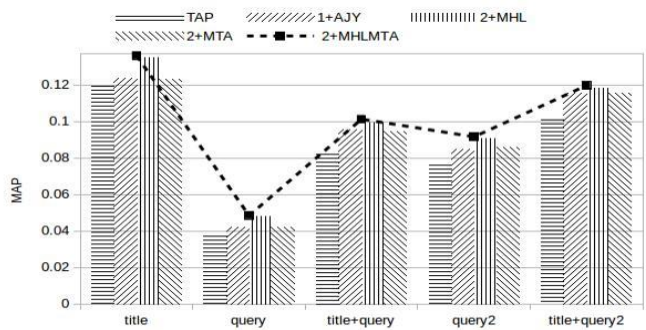

**Figure 1: Graphical comparison of the MAP metric for the 5 different indexing schemes and the 5 different query parsing techniques, when b = 0.75 and k = 1.2.**

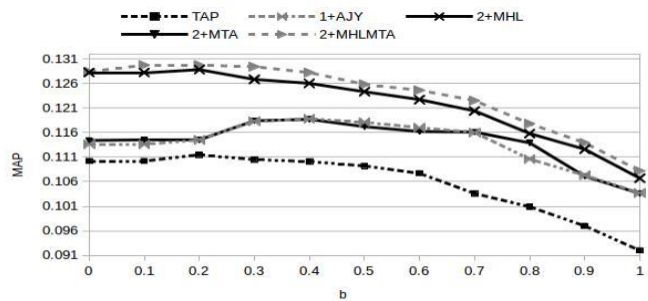

**Figure 2: Graphical comparison of the MAP metric between the 5 different indexing schemata and the 'title&query2' parser, while sweeping the b parameter and k = 1.2.**

I notice from the MAP measurements that the 2+MHLMTA indexing scheme (see section 2.1) performs the best in all 5 query parsing scenarios. Respectively, I observe that the title query parsing technique (see section 2.2) accomplishes the best MAP performance with all 5 indexing scheme scenarios. I can also notice from Figures 2 and 3 that for the query parsing variation title&query2 the highest MAP is achieved by the 2+MHLMTA indexing scheme, when b is 0.3 and k is set to 1.8.

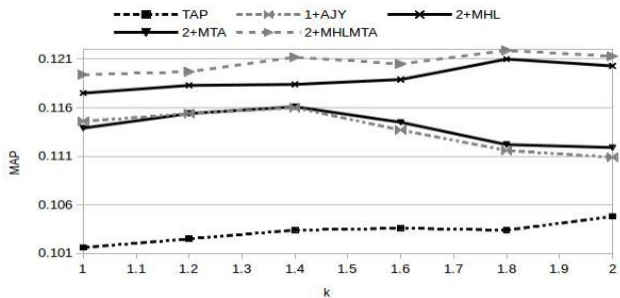

**Figure 3: Graphical comparison of the MAP metric between the 5 different indexing schemes and the 'title&query2' parser, while sweeping the k parameter and b = 0.75.**

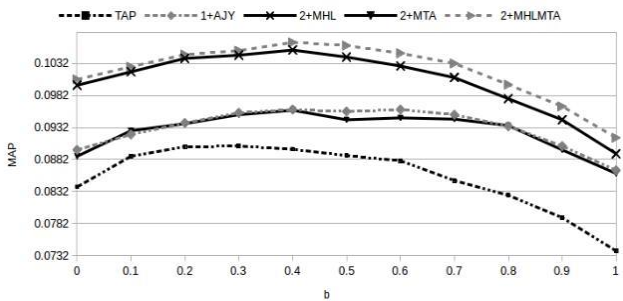

**Figure 4: Graphical comparison of the MAP metric be-tween the 5 different indexing schemes and the 'title&query' parser, while sweeping the b parameter and k = 1.2.**

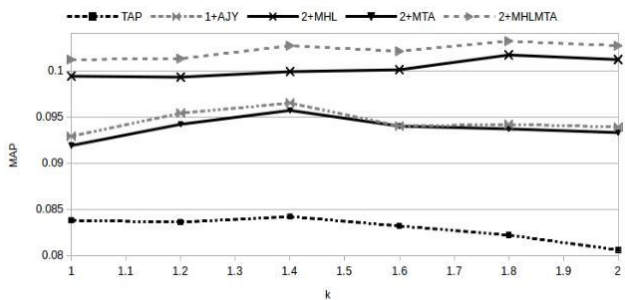

**Figure 5: Graphical comparison of the MAP metric between the 5 different indexing schemata and the 'title&query' parser, while sweeping the k parameter and b = 0.75.**

Correspondingly, from Figures 4 and 5 I can observe that for the query parsing variation title&query the highest MAP is achieved again by the 2+MHLMTA indexing scheme, when b is 0.4 and k is set to 1.8. Furthermore, I noticed from the Precision after 10, 20 and 30 doc-uments retrieved, that there is no dominant indexing methodology.

Moreover, I noticed that the performance of the 1+AJY and 2+MTA indexing methodology (see section 2.1) is very similar in almost every query parsing scenario and for all Precision metrics (precision after 10, 20 and 30 documents). I also observe that the precision after 10, 20 and 30 documents retrieved is higher when the title query parsing technique is utilised instead of the other 4 initial query extraction techniques.

Additionally, I recognise that the measurements of the IRPA at 0.10 and 0.30 recall are always higher when using the 2+MHLMTA indexing scheme. The same can also be said for the IRPA at 0.20 recall, but in this case, there is a small exception when utilising the query topic-parsing technique, where I observed that the 2+MHLMTA indexing scheme accomplishes the best score right after the 2+MHL indexing method. Finally, I observed that the IRPA at 0.10, 0.20 and 0.30 recall (see section 2) is always higher when using the 'title' topic-parsing technique.

## 4 CONCLUSION

I focused on discovering how alternative indexing schemata affect performance and how different extraction techniques of the initial query from CLEF 2017 eHealth Task 2 topics affect performance. I found that performance is improved substantially when using the 2+MHLMTA indexing schema (see section 2.1), which helps me to draw the conclusion that adding more fields to the PubMed indexer is actually a positive contributor for the retrieval effectiveness of an IR system such as L4IR. The query parsing technique 'title' seems to perform the best. However, the encouraging results from the additional query preprocessing of the topic (see section 2.2), which was observed with 'query2' and 'title&query2' gives me great confidence that with elongation of appropriate query preprocessing efforts, the combined usage of Title and Query by an IR system such as L4IR can achieve higher performance than simply using the Title. I look forward to increase the number of my experiments using L4IR, to index additional fields of the PubMed documents and to find new methodologies for indexing and query parsing new empirical test collections. Finally, I plan to test more retrieval models with the L4IR system and make use of Active Learning to enhance the ranking by using the feedback from reviewers.